\documentclass[aps,prl,twocolumn,groupedaddress,showpacs,superscriptaddress,letterpaper,10pt]{revtex4}
\usepackage{graphicx}
\begin{document}

\title{Dynamical Backaction of Microwave Fields on a Nanomechanical Oscillator}

\author{J. D. Teufel}
\email{john.teufel@colorado.edu} \affiliation{JILA,
National Institute of Standards and Technology and the University of
Colorado, Boulder, CO 80309, USA}
\author{J. W. Harlow}
\affiliation{JILA, National Institute of Standards and Technology
and the University of Colorado, Boulder, CO 80309, USA}
\affiliation{Department of Physics, University of Colorado, Boulder,
CO 80309, USA}
\author{C. A. Regal}
\affiliation{Norman Bridge Laboratory of Physics 12-33, California Institute of Technology, Pasadena, California 91125, USA}
\author{K. W. Lehnert}
\affiliation{JILA, National Institute of Standards and Technology
and the University of Colorado, Boulder, CO 80309, USA}
\affiliation{Department of Physics, University of Colorado, Boulder,
CO 80309, USA}

\begin{abstract}
We measure the response and thermal motion of a high-Q nanomechanical oscillator coupled to a superconducting microwave cavity in the resolved-sideband regime where the oscillator's resonance frequency exceeds the cavity's linewidth. The coupling between microwave field and mechanical motion is strong enough for radiation pressure to overwhelm the intrinsic mechanical damping. This radiation pressure damping cools the fundamental mechanical mode by a factor of 5 below the thermal equilibrium temperature in a dilution refrigerator to a phonon occupancy of 140 quanta.
\end{abstract}


\maketitle

Recent advances in coupling mechanical oscillators to electromagnetic resonances have led to accelerated progress in both measuring and cooling mechanical motion \cite{Gigan2006,Arcizet2006,Naik2006,Brown2007,Thomson2008,Regal2008,Kippenberg2008}. In cavity optomechanical systems, the motion of a mechanically compliant object tunes the resonance frequency of an optical cavity, while the intra-cavity light exerts a backaction force on the element \cite{Braginsky1970,Kippenberg2008}. With these systems, observing quantum behavior of mechanical objects seems feasible.  To reach this quantum regime, one must be able to minimize the random thermal motion of the harmonic oscillator so that the zero-point motion becomes dominant.  Furthermore, the oscillator must also be compatible with a measurement technique that is sensitive enough to resolve this residual motion. In the field of cavity optomechanics, the strong interaction between the photons in an electromagnetic cavity and the phonons in a mechanical oscillator provides both sensitive measurement and the potential for ground-state cooling.

Despite the large optomechanical coupling achieved with high-finesse Fabry-Perot cavities, mechanical ground-state cooling remains elusive. For optical backaction to cool a mechanical oscillator to its ground state, the cavity optomechanical system must reach the resolved-sideband limit where the oscillator's resonance frequency $\omega_m$ exceeds the cavity linewidth $\gamma$.  (This condition is analogous to the requirements for laser sideband cooling first demonstrated with trapped ions several decades ago \cite{Diedrich1989}.) In cavity optomechanics, only the highest-finesse cavities approach this limit. It is only in recent experiments with a microtoroid simultaneously serving as both the mechanical and optical 
resonator that $\omega_m > \gamma$ is achieved \cite{Schliesser2008}. While reaching the ground state of the oscillator is aided by precooling in a cryostat, this is technically challenging in optical systems \cite{Groblacher2008}.

\begin{figure}
\includegraphics{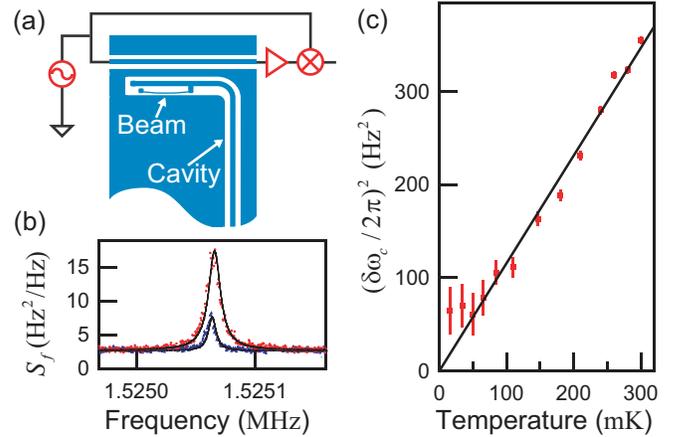}\\
\caption{\label{fig1} (color online). (a) Sketch of a nanomechanical beam embedded in a microwave cavity with a measurement schematic. The solid (blue) region represents the patterned aluminum film. The cavity is excited by a microwave signal propagating in a nearby transmission line. The motion of the oscillator is encoded in the amplitude and phase of the transmitted signal. We recover the motion of the oscillator by amplifying and mixing this signal with a reference. (b) Power spectral density of cavity frequency fluctuations ($f=\omega_c/2\pi$). Data are shown for two different cryostat temperatures of 50~mK (lower) and 240~mK (upper) with a Lorentzian fit (solid lines) to each trace. (c) The mean-square fluctuations in the cavity's resonance frequency due to the thermal motion of the oscillator as a function of cryostat temperature.  The linear fit (solid line) yields $g=2\pi \times 6.4~$kHz/nm.}
\end{figure}

In this Letter, we demonstrate a microwave optomechanical system that is well into the resolved-sideband limit ($\omega_m = 6.6\times \gamma$) and is compatible with precooling in a dilution refrigerator.  Within a superconducting, microwave, transmission-line resonator, we embed a nanomechanical oscillator in the form of a freely suspended, doubly clamped beam [Fig.~\ref{fig1}(a)].  We show that the frequency and power of the microwave signal applied to the cavity strongly affect the mechanical properties of the oscillator.  These effects are in excellent agreement with the theoretical predictions for the dynamical backaction in the resolved-side band regime.  We also measure the thermal motion of the beam and show that the small number of thermal quanta ($\bar{m} = 700$) of the mechanical mode in equilibrium with a 50~mK dilution refrigerator can be further cooled using dynamical backaction to $\bar{m} = 140$.  These measurements demonstrate that the mechanical damping can be dominated by the radiation pressure effects, indicating that this system is an excellent candidate for probing the quantum behavior of a tangible harmonic oscillator.

Unlike traditional optomechanical systems where the mechanically compliant object must be large enough to focus a laser on its surface, microwave cavities can efficiently couple to mechanical oscillators whose cross-sectional area is limited only by the lithographic techniques used to fabricate them. Lighter oscillators respond to the smaller radiation pressure forces imparted by microwave photons because they have larger zero-point motion, $x_{zp}=\sqrt{\hbar/(2 m \omega_m)}$, where $m$ is the effective mass of the oscillator and $x$ is the displacement at the center of the beam. Together, the narrow linewidth of a superconducting microwave cavity  and the large mechanical resonance frequency of a light oscillator easily achieve the resolved-sideband limit.  Finally, the resulting microwave, optomechanical system is naturally compatible with low temperature ($T<1$~K) operation \cite{Regal2008,Teufel2008}. The low thermal occupancy of the mechanical mode greatly reduces the radiation pressure effects necessary to cool to the ground state.

Recent theory \cite{Marquardt2007,Wilson2007} predicts that radiation pressure forces at frequencies near $\omega_m$ alter the mechanical oscillator's response function. When the cavity is excited with a tone at frequency $\omega_e$, this dynamical backaction adds a term $\Gamma$ to the oscillator's damping rate and $\Omega$ to its resonance frequency, where
\begin{equation}\label{gamma}
\Gamma=B\left[\frac{\gamma}{\gamma^{2}+4\left(\Delta+\omega_{m}\right)^{2}}-\frac{\gamma}{\gamma^{2}+4\left(\Delta-\omega_{m}\right)^{2}}\right],\\
\end{equation}
\begin{equation}\label{omega}
\Omega =B\left[\frac{\Delta+\omega_{m}}{\gamma^{2}+4\left(\Delta+\omega_{m}\right)^{2}}+\frac{\Delta-\omega_{m}}{\gamma^{2}+4\left(\Delta-\omega_{m}\right)^{2}}\right],
\end{equation}
and $\Delta=\omega_e-\omega_c$ is the detuning between the frequency of excitation and the cavity's resonance.  The prefactor $B=4\bar{n} g^{2} x_{zp}^{2}$ controls the strength of the optomechanical interaction. Here $\bar{n}$ is the average number of photons in the cavity and $g$ is the optomechanical coupling ($d\omega_c/dx$). When the microwave signal is applied above the cavity resonance ($\Delta > 0$), $\Gamma < 0$  and the oscillator's motion is amplified. When the total damping, the sum of the intrinsic dissipation $\gamma_{m0}$ and the radiation damping, becomes negative ($\gamma_m=\gamma_{m0}+\Gamma<0$), the mechanical motion undergoes regenerative oscillations \cite{Braginsky2001}. By detuning below the cavity resonance, the increased damping is accompanied by cooling of the mechanical motion. In the presence of radiation damping, the oscillator is coupled to two baths: one characterized by the physical temperature of its surroundings $T_0$, and the other by an effective nonequilibrium temperature of the radiation pressure force $T_p$.  The final temperature is given by the weighted average of these two quantities: $T_m=(\gamma_{m0} T_0 + \Gamma T_p)/\gamma_m$ \cite{Marquardt2007,Wilson2007}.

In our device, the cavity is a quarter-wave coplanar waveguide (CPW) resonator \cite{Day2003} with $\omega_c = 2\pi \times 5.22$~GHz, and  $\gamma = 2\pi \times 230$~kHz.  The mechanical oscillator is a doubly clamped beam with dimensions of $100 \times 0.13 \times 0.12$~$\mu$m$^{3}$ and has an effective mass of $6.2 \times 10^{-15}$~kg. The beam is freely suspended within a voltage antinode of the cavity, and its in-plane motion capacitively loads the cavity and modulates the electrical length.  These devices are fabricated by patterning a resist mask with a combination of optical and electron-beam lithography, and then depositing both the microwave cavities and mechanical oscillators with a single evaporation of aluminum [Fig.~\ref{fig1}(a)]. The device is annealed at $340^{\circ}$C for 30 minutes, and then the beam is suspended using an SF$_{6}$ reactive-ion etch to remove the silicon substrate in a small region around the nanomechanical beam. The annealing step increases the tensile stress in the aluminum film which increases both $\omega_m$ and the quality factor ($Q_m=\omega_m/\gamma_m$) by 50 times what is measured in unstressed devices \cite{Regal2008,Teufel2008}. At low temperatures, $\omega_m= 2\pi \times 1.525$~MHz and $Q_m > 300,000$. All measurements of the device are made in a dilution refrigerator using homodyne detection shown schematically in Fig.~\ref{fig1}(a) and discussed in previous work \cite{Regal2008}.

\begin{figure}
\includegraphics{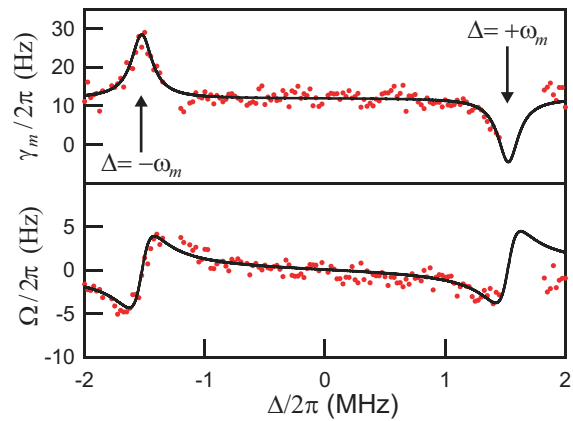}\\
\caption{\label{fig2} (color online). The mechanical damping and resonance frequency shift as functions of detuning.  As the detuning is changed, the incident microwave power is adjusted to keep the circulating power constant. The solid lines are a coupled fit to Eqs.\ (\ref{gamma}) and (\ref{omega}) where the only free parameter is the circulating microwave power.  This power agrees within the 3~dB uncertainty with which the attenuation within the cryostat is known. Data is missing near $\Delta=\omega_m$ where the beam regeneratively oscillated, as expected for $\gamma_m<0$.}
\end{figure}

We infer the displacement of the nanomechanical beam by monitoring the amplitude and phase fluctuations of the microwave field emerging from the cavity. We can also apply an electrostatic force to the beam by varying the potential of a nearby electrode \cite{Regal2008}. To characterize the optomechanical coupling, we use the thermal mechanical motion as a calibration [Fig.~\ref{fig1}(b) and (c)]. Even in the absence of any applied force, the measurement easily resolves a peak in the power spectral density of the demodulated microwave signal at the mechanical resonance frequency. This peak is due to the thermal motion of the beam; therefore, the area of the peak should be proportional to temperature in accordance with the equipartition theorem. We indeed find that above 50~mK the peak area follows a linear relationship with the cryostat temperature, from which we find an optomechanical coupling of $g=2\pi \times 6.4$~kHz/nm. In the resolved-sideband limit, the signal-to-noise ratio of the measurement is best when the microwave excitation is detuned from the cavity resonance by the mechanical resonance frequency.  As this is also the point of strongest dynamical backaction, we perform this calibration with $\Delta=-\omega_m$, but at a very low circulating power of $P_{c}=\hbar \omega_e^{2} \bar{n}=50$~nW. At this low power, there is no discernable difference in the mechanical properties between detuning to either side of the cavity resonance.

As the microwave power is increased, the effects of the radiation as given by Eqs. (\ref{gamma}) and (\ref{omega}) become apparent.  Figure \ref{fig2} shows the mechanical linewidth and resonance frequency as functions of detuning while holding $P_c$ constant. At each detuning, we measure the mechanical response to an applied force in order to extract $\omega_m$ and $\gamma_m$. Note that in this resolved-sideband limit, the change in the beam's mechanical properties with detuning around $\Delta=\pm \omega_m$ reproduces the complex response of the cavity. 
The agreement with Eqs.\ (\ref{gamma}) and (\ref{omega}) demonstrates that it is the finite response time of the cavity that is responsible for the damping.  The independently determined values of $\omega_m$, $\gamma$, and $g$ accurately account for the location, width, and height of the features in Fig.~\ref{fig2}. For this intermediate circulating power of 900~nW, the magnitude of the radiation damping is slightly greater than the intrinsic damping ($\Gamma\gtrsim \gamma_{m0}$).  The mechanical linewidth is roughly doubled for $\Delta=-\omega_m$ and becomes negative near $\Delta=+\omega_m$.  In this region of negative damping, we observe regenerative oscillations of the beam.  

\begin{figure}
\includegraphics{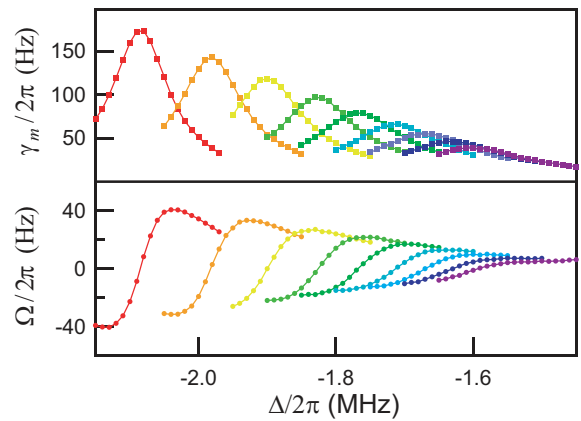}\\
\caption{\label{fig3} (color online).  The mechanical damping and resonance frequency shift as functions of detuning for microwave powers that drive the cavity nonlinear.  Each curve corresponds to a different incident microwave power, and the detuning is relative to the low-power resonance frequency of the cavity. As the incident power is increased from 16~nW (violet, right) to 100~nW (red, left) in 1 dB steps, the absolute frequency at which maximum mechanical damping occurs follows the nonlinear cavity resonance to lower frequency.}
\end{figure}

We next increase the microwave power in order to understand how large these radiation effects can be.  There is a practical limit on $P_c$ in our cavity set by the strength of the superconductivity in the aluminum that defines the cavity.  Beyond $P_{c}\approx 1$~$\mu$W, the cavity resonance becomes nonlinear and eventually bistable.  This effect does not preclude mechanical measurements; however, in this regime, changing the microwave power changes the cavity lineshape and shifts its resonance frequency. We measure the mechanical response and find $\gamma_m$ and $\omega_m$ at the largest microwave powers (Fig.~\ref{fig3}).  To avoid regenerative mechanical oscillations, we only measure in the region $\Delta<0$.  Because the detuning is referenced to the low-power resonance frequency of the cavity, each higher power curve has an optimum detuning (maximal $\gamma_m$) that is lower in frequency.  This is a direct measure of the nonlinear shift in the cavity's resonance frequency.  Unlike the data in Fig.~\ref{fig2}, these curves are taken with a constant \textit{incident} power $P_i$. For our geometry in the overcoupled regime, $P_i=\hbar \omega_e \bar{n}(\gamma^2+4\Delta^2)/\gamma$. These data demonstrate a microwave optomechanical system operating both in the resolved-sideband and the radiation-dominated limit ($\Gamma\gg\gamma_{m0}$).

\begin{figure}
\includegraphics{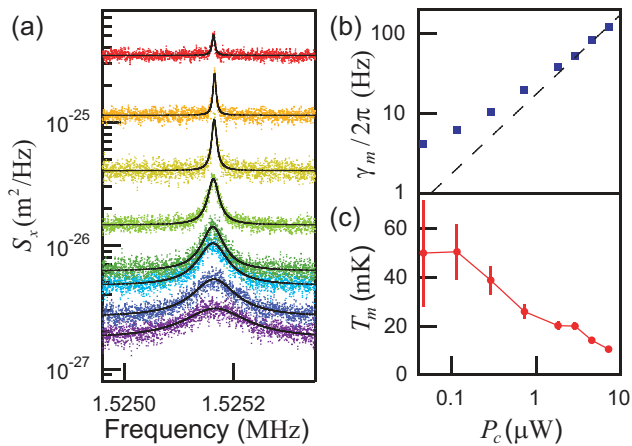}\\
\caption{\label{fig4} (color online) (a) Thermal noise of the mechanical motion.  Each noise spectrum (dots) is taken at a new microwave power such that the microwave carrier is optimally detuned from the cavity resonance for maximum mechanical damping and cooling.  The solid lines show Lorentzian fits, which are used to determine $\gamma_m$ and $T_m$. The data are not offset; as the circulating power is increased from 46~pW (upper, red) to 7.3~$\mu$W (lower, violet), the absolute measurement imprecision improves. (b) The total damping (squares) and a dashed line indicating $\gamma_m \propto P_{c}$ as predicted by Eq. (1) when $\Gamma\gg\gamma_{m0}$. (c) The temperature of the mechanical mode, as determined from the area of the noise spectra, shows cooling from 50~mK to 10~mK.}
\end{figure}

Since the unique benefit of operating in the resolved-sideband limit is the potential to cool the mechanical mode of interest to the ground state \cite{Marquardt2007,Wilson2007,Dobrindt2008}, we now examine the oscillator's temperature as a function of microwave power.  To do this, the detuning is adjusted to the optimal cooling point, as determined from the data in Fig.~\ref{fig3}.  Because we are using the microwave signal both to cool the beam and to measure the displacement, it is important that we have a valid calibration.  We apply a constant electrostatic force to the beam whose magnitude we know from our low-power thermal calibration (Fig.~\ref{fig1}).  In this way, the measured power spectral density can be converted to displacement units, as shown in Fig.~\ref{fig4}(a). Because of the stronger optomechanical coupling and higher mechanical resonance frequency, the displacement sensitivity is not limited by the cavity frequency noise as seen in previous experiments \cite{Regal2008}; the absolute sensitivity improves with increasing power even at the highest microwave power. Furthermore, as the microwave power is increased, the thermal peak becomes wider (lower $Q_m$) and smaller (less area), indicating that the additional damping is accompanied by cooling.  Because $\omega_m = 6.6\times \gamma$, the effective temperature of the photon bath is negligible ($T_p\approx 10$~$\mu$K) compared to that of the cryogenic environment \cite{Marquardt2007,Wilson2007}.  Thus, the temperature of the mechanical mode in the presence of radiation effects should be determined from the increase in damping, $T_{m} \approx T_0(\gamma_{m0}/\gamma_m)$. Figure~\ref{fig4}(b) shows that at the highest powers the damping is increased by a factor of 30 and the temperature is reduced by a factor of 5.  We suspect this discrepancy arises from the fact that large microwave power may heat the local environment \cite{Regal2008}.  Because $\gamma_{m0}$ increases with temperature, parasitic heating would reduce our cooling in two ways, both by warming the bath to which the beam is coupled and by coupling more strongly to that bath.  Nevertheless, Fig.~\ref{fig4}(b) shows that at the highest powers, the damping becomes linear in power as expected from radiation pressure effects [Eq. (1)], overwhelming the weakly power dependent intrinsic damping $\gamma_{m0}$ and cooling the beam to 10~mK. The natural metric to quantify this cooling is how many thermal quanta remain in the mechanical oscillator.  For this device, cooling to 10~mK corresponds to a final phonon occupation of $\bar{m}=140 \pm 7$.  This cooling is currently limited by the microwave power that can be applied to the cavity, not by a fundamental limitation of the dynamical backaction.  For example, at the point of maximum cooling, the mechanical oscillator still has a quality factor of approximately 10,000.

In future devices, several improvements should allow us to exploit the full-cooling potential of the resolved-sideband regime.  Fabricating the cavity out of a higher transition temperature superconducting metal, such as niobium, will greatly increase the circulating power at which the resonance becomes nonlinear. At higher power, the cavity frequency noise creates a larger radiation-pressure force on the beam, but with a modest increase in the coupling strength, this random force will not prevent $\bar{m}<1$. Once cooling at or near the ground state becomes technically feasible, an equally important component is a measurement that is sensitive enough to resolve the residual quantum motion.  To this end, we will integrate this measurement technique with a quantum-limited microwave amplifier that will allow displacement measurements with improved sensitivity \cite{Castellanos2008}.

In conclusion, we have shown that by coupling a superconducting microwave resonator to the motion of a nanomechanical oscillator, one can reach the resolved-sideband limit.  The coupling is strong enough that the dynamical backaction of the microwave photons can become the dominant source of mechanical dissipation, a prerequisite for cooling to the mechanical ground state.  This damping cools the fundamental mechanical mode of the oscillator fivefold to a thermal occupation of 140 phonons.  These measurements indicate that this optomechanical system is a promising candidate for realizing ground-state cooling of a mechanical oscillator.

\begin{acknowledgments}
The authors acknowledge funding from the National Institute of Standards and Technology (NIST) and from the National Science Foundation; C.A.R.\ acknowledges support from the Millikan Postdoctoral Fellowship.  We thank M.\ A.\ Castellanos-Beltran and N.\ E.\ Flowers-Jacobs for discussions and technical assistance.
\end{acknowledgments}


\begin{thebibliography}{10}

\bibitem{Gigan2006}
S. Gigan {\it et~al.}, Nature {\bf 444},  67  (2006).

\bibitem{Arcizet2006}
O. Arcizet {\it et~al.}, Nature {\bf 444},  71  (2006).

\bibitem{Naik2006}
A. Naik {\it et~al.}, Nature {\bf 443},  193  (2006).

\bibitem{Brown2007}
K.~R. Brown {\it et~al.}, Phys. Rev. Lett. {\bf 99},  137205  (2007).

\bibitem{Thomson2008}
J.~D. Thompson {\it et~al.}, Nature {\bf 452},  72  (2008).

\bibitem{Regal2008}
C.~A. Regal, J.~D. Teufel, and K.~W. Lehnert, Nature Phys. {\bf 4}, 555 (2008).

\bibitem{Kippenberg2008}
T.~J. Kippenberg and K.~J. Vahala, Science {\bf 321}, 1172 (2008) and references therein.

\bibitem{Braginsky1970}
V.~B. Braginsky, A.~B. Manukin, and M.~Y. Tikhonov, Sov. Phys. JETP {\bf 31},
  829  (1970).

\bibitem{Diedrich1989}
F. Diedrich {\it et~al.}, Phys. Rev. Lett. {\bf 62},  403  (1989).

\bibitem{Schliesser2008}
A. Schliesser {\it et~al.}, Nature Phys. {\bf 4},  415  (2008).

\bibitem{Groblacher2008}
S. Gr\"{o}blacher {\it et~al.}, Europhys. Lett. {\bf 81},  54003  (2008).

\bibitem{Teufel2008}
J.~D. Teufel, C.~A. Regal, and K.~W. Lehnert, New J. Phys. {\bf 10} 095002 (2008).

\bibitem{Marquardt2007}
F. Marquardt {\it et~al.}, Phys. Rev. Lett. {\bf 99},  093902  (2007).

\bibitem{Wilson2007}
I. Wilson-Rae {\it et~al.}, Phys. Rev. Lett. {\bf 99},  093901  (2007).

\bibitem{Braginsky2001}
V.~B. Braginsky, S.~E. Strigin and S.~P. Vyatchanin, Phys. Lett. A {\bf 287}, 331 (2001).

\bibitem{Day2003}
P.~K. Day {\it et~al.}, Nature {\bf 425},  817  (2003).

\bibitem{Dobrindt2008}
J.M. Dobrindt, I. Wilson-Rae, and T.J. Kippenberg, arXiv:0805.2528v1.

\bibitem{Castellanos2008}
M.~A. Castellanos-Beltran {\it et~al.}, Nature Phys. (to be published) doi:10.1038/nphys1090.

\end{thebibliography}
\end{document}